\documentstyle[aps,leqno,preprint]{revtex}  
\input{epsf}
\begin{document}
\renewcommand{\thefootnote}{\alph{footnote} }
\narrowtext
%
%
\hoffset=0.0in
%
%
\def\DZ{\ifmmode D^0 \else $D^0$\fi}
\def\DZb{\ifmmode \bar{D}^0 \else $\bar{D}^0$\fi}
\def\DP{\ifmmode D^+ \else $D^+$\fi}
\def\DSP{\ifmmode D^{*+} \else $D^{*+}$\fi}
\def\DSM{\ifmmode D^{*+} \else $D^{*-}$\fi}
\def\DST{\ifmmode D^* \else $D^*$\fi}
\def\DM{\ifmmode D^- \else $D^-$\fi}
\def\DPM{\ifmmode D^\pm \else $D^\pm$\fi}
\def\Mmin{\ifmmode M_{\rm min} \else $M_{\rm min}$\fi}
\def\PT{\ifmmode p_T \else $p_T$\fi}
\def\KM{\ifmmode K^- \else $K^-$\fi}
\def\kstar{\ifmmode K^{*0} \else $K^{*0}$\fi}
\def\kstarb{\ifmmode \overline{K}^{*0} \else $\overline{K}^{*0}$\fi}
\def\kstarbno{\ifmmode \overline{K}^{*} \else $\overline{K}^{*}$\fi}
\def\kst892{\ifmmode \overline{K}^{*}(892)^0 \else
$\overline{K}^{*}(892)^0$\fi}
\def\kstz{\ifmmode \overline{K}^{*}(892)^{0}
  \else $\overline{K}^{*}(892)^{0}$\fi}
\def\KSTM{\ifmmode \overline{K}^{*-} \else $\overline{K}^{*-}$\fi}
\def\kstm890{\ifmmode \overline{K}^{*}(892)^-
\else $\overline{K}^{*}(892)^-$\fi}
\def\kst{\ifmmode \overline{K}^{*}(892) \else
$\overline{K}^{*}(892)$\fi}
\def\rh{\ifmmode \rho(770) \else $\rho(770)$\fi}
\def\rh0{\ifmmode \rho(770)^{0} \else $\rho(770)^{0}$\fi}
\def\rhm{\ifmmode \rho(770)^{-} \else $\rho(770)^{-}$\fi}
\def\munu{\ifmmode \mu^{+}\nu \else $\mu^{+}\nu$\fi}
\def\muX{\ifmmode \mu^{+}X \else $\mu^{+}X$\fi}
\def\lnu{\ifmmode \ell\nu \else $\ell\nu$\fi}
\def\nmu{\ifmmode \nu_\mu \else $\nu_\mu$\fi}
\def\nmub{\ifmmode \bar{\nu}_\mu \else $\bar{\nu}_\mu$\fi}
\def\pmu{\ifmmode p_\mu \else $p_\mu$\fi}
\def\ptmu{\ifmmode p_{T\mu} \else $p_{T\mu}$\fi}
\def\pth{\ifmmode p_{T h} \else $p_{T h}$\fi}
\def\bmu{\ifmmode b_\mu \else $b_\mu$\fi}
\def\pperp{\ifmmode p_\perp \else $p_\perp$\fi}
\def\xf{\ifmmode x_F \else $x_F$\fi}
\def\pts{\ifmmode {p_T}^2 \else ${p_T}^2$\fi}
\def\pt{\ifmmode {p_T} \else ${p_T}$\fi}
\def\chisq{\ifmmode \chi^2 \else $\chi^2$\fi}

\def\pvec{\ifmmode \underline{p} \else $\underline{p}$\fi}
%
\def\sameauthors#1
{\hbox to\textwidth{\hss\vrule height.3cm width0pt\relax%
#1\hss}}
\def\etal{{\it et al.}\rm}
%

\title{
{\normalsize
\begin{flushright}
\renewcommand{\arraystretch}{0.8}
\tabcolsep=0.0mm
\begin{tabular}{l}
Phys.\ Lett.\ B {\bf{403}} (1997) 185\\
UMS/HEP/96-001 \\
FERMILAB-Pub-96-206-E\\
                     \\
\end{tabular}
\end{flushright}} 
Observation of $D-\pi$ Production Correlations \\    
in 500 GeV $\pi^--N$ Interactions} 

\author{
    E.~M.~Aitala,$^8$
       S.~Amato,$^1$
    J.~C.~Anjos,$^1$
    J.~A.~Appel,$^5$
       D.~Ashery,$^{14}$
       S.~Banerjee,$^5$
       I.~Bediaga,$^1$
       G.~Blaylock,$^2$
    S.~B.~Bracker,$^{15}$
    P.~R.~Burchat,$^{13}$
    R.~A.~Burnstein,$^6$
       T.~Carter,$^5$
 H.~S.~Carvalho,$^{1}$
 N.~K.~Copty,$^{12}$ 
       I.~Costa,$^1$
    L.~M.~Cremaldi,$^8$
 C.~Darling,$^{18}$
       K.~Denisenko,$^5$
       A.~Fernandez,$^{11}$
       P.~Gagnon,$^2$
       S.~Gerzon,$^{14}$
       K.~Gounder,$^8$
     A.~M.~Halling,$^5$
       G.~Herrera,$^4$
 G.~Hurvits,$^{14}$
       C.~James,$^5$
    P.~A.~Kasper,$^6$
       S.~Kwan,$^5$
    D.~C.~Langs,$^{10}$
       J.~Leslie,$^2$
       B.~Lundberg,$^5$
       S.~MayTal-Beck,$^{14}$
       B.~T.~Meadows,$^3$
 J.~R.~T.~de~Mello~Neto,$^1$
    R.~H.~Milburn,$^{16}$
 J.~M.~de~Miranda,$^1$
       A.~Napier,$^{16}$
       A.~Nguyen,$^7$
  A.~B.~d'Oliveira,$^{3,11}$
       K.~O'Shaughnessy,$^2$
    K.~C.~Peng,$^6$
    L.~P.~Perera,$^3$
    M.~V.~Purohit,$^{12}$
       B.~Quinn,$^8$
       S.~Radeztsky,$^{17}$
       A.~Rafatian,$^8$
    N.~W.~Reay,$^7$
    J.~J.~Reidy,$^8$
    A.~C.~dos Reis,$^1$
    H.~A.~Rubin,$^6$
 A.~K.~S.~Santha,$^3$
 A.~F.~S.~Santoro,$^1$
       A.~J.~Schwartz,$^{10}$
       M.~Sheaff,$^{17}$
    R.~A.~Sidwell,$^7$
    A.~J.~Slaughter,$^{18}$
    M.~D.~Sokoloff,$^3$
       N.~R.~Stanton,$^7$
       K.~Stenson,$^{17}$  
    D.~J.~Summers,$^8$
 S.~Takach,$^{18}$
       K.~Thorne,$^5$
    A.~K.~Tripathi,$^9$
       S.~Watanabe,$^{17}$
 R.~Weiss-Babai,$^{14}$
       J.~Wiener,$^{10}$
       N.~Witchey,$^7$
       E.~Wolin,$^{18}$
       D.~Yi,$^8$
       S.~Yoshida,$^7$
       R.~Zaliznyak,$^{13}$
       C.~Zhang$^7$ \\
\begin{center} (Fermilab E791 Collaboration)\end{center}
}
\address{
$^1$ Centro Brasileiro de Pesquisas F\'\i sicas, Rio de Janeiro, Brazil\\
$^2$ University of California, Santa Cruz, California 95064\\
$^3$ University of Cincinnati, Cincinnati, Ohio 45221\\
$^4$ CINVESTAV, Mexico\\
$^5$ Fermilab, Batavia, Illinois 60510\\
$^6$ Illinois Institute of Technology, Chicago, Illinois 60616\\
$^7$ Kansas State University, Manhattan, Kansas 66506\\
$^8$ University of Mississippi, University, Mississippi 38677\\
$^9$ The Ohio State University, Columbus, Ohio 43210\\
$^{10}$ Princeton University, Princeton, New Jersey 08544\\
$^{11}$ Universidad Autonoma de Puebla, Mexico\\
$^{12}$ University of South Carolina, Columbia, South Carolina 29208\\
$^{13}$ Stanford University, Stanford, California 94305\\
$^{14}$ Tel Aviv University, Tel Aviv, Israel\\
$^{15}$ 317 Belsize Drive, Toronto, Canada\\
$^{16}$ Tufts University, Medford, Massachusetts 02155\\
$^{17}$ University of Wisconsin, Madison, Wisconsin 53706\\
$^{18}$ Yale University, New Haven, Connecticut 06511\\
}

\date{19 June 1997}

\maketitle

\begin{abstract}

We study the charge correlations between charm mesons 
produced in  500 GeV $\pi^--N$ interactions and the charged pions
produced closest to them in phase space. With 110,000 fully 
reconstructed {\em D} mesons from experiment E791 at Fermilab, the 
correlations are studied as functions of the  $ D\pi - D $ mass difference 
and of Feynman $ x $. 
We observe significant correlations which appear to originate
from a combination of sources including fragmentation
dynamics, resonant decays, and charge of the beam. \hfill\break
\end{abstract}


While the production of heavy quarks can be calculated in perturbative Quantum
Chromodynamics (QCD), the evolution of these quarks into hadrons remains one of
the most challenging aspects of nonperturbative QCD.  Correlations between
charm mesons and the charged pions produced closest to them in phase space
provide information on how quarks evolve into hadrons.  
Fragmentation dynamics\cite{the:ros2},
resonances\cite{the:ros1},
and beam effects can each produce such correlations.  The
relative importance of these mechanisms must be determined from data.

During fragmentation, correlations could be produced because $ \overline q
q $ pairs from the vacuum are neutral.
For example, if a $ c $ quark combines with a $ \overline d $  from
such a pair to form a $ D^+ $, the remaining $ d $
is close by in phase space and is likely to become part of the
closest pion, which we call the ``associated pion''.
Thus, $ D^+ \pi^- $ ($D^- \pi^+$) would be favored and 
$ D^+ \pi^+ $ ($D^- \pi^- $) disfavored.
Similarly, $ D^0 \pi^+ $ ($ \overline {D {}^0} \pi^- $) would be favored 
and $ D^0 \pi^- $ ($ \overline {D {}^0}\pi^+ $) disfavored.
Resonances produce the same favored associations.
$ D^{*+} $ decay associates a $ \pi^+ $ with a $ D^0 $ while
$ D^{*-} $ decay associates a $ \pi^- $ with a $\overline {D {}^0} $.
Qualitatively, $ D^{**} $ decays produce the same correlations. 

The charge of the beam particle can also lead to charge correlations.
Using a $ \pi^- $ beam can lead to the association of both charm mesons 
and anticharm mesons with negative pions, especially in the
forward (beam) direction.
Two distinct but related mechanisms can lead to this result.
If the charm quark (antiquark) produced in a hard interaction
coalesces with the antiquark (quark)
from the beam particle to form the charm (anticharm) meson, the 
remaining quark (antiquark) from the beam can become part of
a negative pion, but not part of a positive pion.
If neither the quark nor the antiquark from the beam pion is
used in making the charm meson, both are available to
form negative pions but not positive pions.

By comparing the charge correlations of different species of charm mesons
and antimesons with associated pions, and by studying them as functions
of Feynman $ x $ ($x_F$) of the charm meson, one can hope to disentangle 
some of these processes. Evidence of such correlations between $ B $ mesons 
and associated light mesons, ascribed to resonances, has been observed
in $ Z^0 $ decays at LEP by the OPAL
collaboration\cite{exp:opal}.
In this letter, we report the first observation of fragmentation and
beam-related production 
correlations for charm mesons.

We use $D^0 \to K^- 
\pi^+$, $D^+ \to K^- \pi^+ \pi^+$, and $D^{*+} \to D^0 \pi^+$ signals
(and their charge conjugate decays)
from experiment E791 at Fermilab for this study.
The data were recorded using a 500 GeV/c $\pi^-$ beam 
interacting in five thin target foils (one platinum, four diamond) separated 
by gaps of about 1.4 cm. The detector, described 
elsewhere in more detail\cite{exp:det}, is a large-acceptance, forward, 
two-magnet spectrometer.
Its key components for this study include eight planes of multiwire 
proportional chambers, six
planes of silicon microstrip detectors (SMD) before the target for 
beam tracking,
a 17-plane~SMD system and 35~drift chamber planes downstream of the target  
for track and vertex reconstruction,
and two multicell threshold {\v C}erenkov counters for charged particle
identification.

During event reconstruction, all events with evidence of well-separated 
production and decay vertices were retained as charm decay candidates.
For this study, we require the candidate charm decay vertex 
to be located well outside the target foils and to be at least 
8$\sigma_{\Delta}$ downstream 
of the primary vertex (where $\sigma_{\Delta}$ is the error 
in the measured longitudinal separation between the vertices 
$\approx 350\mu$m). 
The momentum vector of the candidate 
{\em D} must point back to the primary vertex 
with impact parameter less than 80$\mu$m. 
The momentum of the {\em D} 
transverse to the line joining the primary and secondary vertices 
must be less than 0.35 GeV/$c$. Each 
decay track must pass closer to the secondary vertex than to the 
primary vertex. 
Finally, the track assigned to be the kaon in the charm decay must 
have a signature in the 
{\v C}erenkov counters consistent 
with the kaon hypothesis. 
The $D^{*\pm}$ candidates  
are found from the $D^0/\overline {D {}^0}$ samples by
adding $\pi^\pm$ tracks and requiring that the mass difference
$\Delta m$ =  $M(D\pi)-M(D)$
be consistent with the $ D^* \to D \pi $ hypothesis.
The final signal sizes are obtained by fitting 
the invariant mass spectra as Gaussian signals and linear backgrounds. For
$D^0$, $\overline{D {}^0}$, $D^+$, $D^-$, $D^{*+}$, and $D^{*-}$, 
the fits yield 
$22587\pm210$, $24237\pm216$, $24569\pm204$, $29649\pm238$, $4997\pm84$ and 
$6048\pm93$ events, respectively. The r.m.s. mass resolutions, $\sigma_{D}$,
used later in defining signal and background bands, 
are $13~{\rm MeV}/c^2$, $13~{\rm MeV}/c^2$, and $14~{\rm MeV}/c^2$ for
$D^0$, $D^+$, and $D^{*+}$, respectively. 

For each {\em D} found in an event, all tracks originating from the
primary vertex and producing a pion signature in the {\v C}erenkov counters
are combined with the {\em D}.
Among these combinations, the pion that forms the
smallest invariant mass difference ($\Delta m_{min}$)
with the {\em D} decay products is selected as the associated
pion. 

We define the correlation parameter $\alpha$ as
$$ \alpha(D) \equiv \frac{\sum N_i(D\pi^q)-\sum N_i(D\pi^{-q})}
 {\sum N_i(D\pi^q)+\sum N_i(D\pi^{-q})}, \eqno(1)$$
\noindent where $ q $ =  +, $ - $, $ - $, +, $ - $, + for $D = D^0$, 
$\overline {D {}^0}$ , $ D^+$, $ D^-$,
$D^{*+}$, and $D^{*-}$, respectively, and $\sum N_i(D\pi^q)$ 
denotes the number of charm mesons for which the selected pion has the charge
$ q $.
In the absence of correlations
$\alpha$ is zero, and in maximally correlated cases it is unity. 

We first study the $D\pi$ correlations as functions of 
$\Delta m_{min}$ for $\Delta m_{min} <0.74$ GeV/$c^2$. 
The number of $D\pi$ signal combinations in each $\Delta m_{min}$ bin
is determined by subtracting from the
$\Delta m_{min}$ distribution for $D$ candidates (mass within 
$\pm2.5~\sigma_D$ of the nominal $D$ mass) the appropriately normalized
$\Delta m_{min}$ distribution for background events (mass between
3.0 $\sigma_D$ and 5.5 $\sigma_D$ from the nominal $D$ mass).
The correlation parameters for 
background-subtracted signals (before additional corrections)
and background regions are listed in Table 1. The signal correlations
differ significantly from the background correlations. We note 
that replacing the $D$ candidate in an event with a $D$ of the 
same species from another event, while keeping the rest of the event the same,
produces correlations consistent with those of the background.

We use a Monte Carlo simulation of the experiment and
the LUND event generator (PYTHIA 5.7/JETSET 7.3)\cite{the:lmc} 
to model the effects of our apparatus and reconstruction.
This simulation describes the geometry, resolution, noise, and efficiency of 
all detectors, as well as interactions and decays in the spectrometer. 
The detected $D^*/D$ production ratio in the Monte Carlo matches our data well.  
As with real events, the associated pion for each
reconstructed {\em D} meson is selected. By matching the selected
pion's momentum vector with the momenta of all generated particles,
we determine whether
the selected pion track is a real track or a ghost (false) track. 
Selecting a ghost pion (not matched to any generated track) or a real pion 
not matched to the true associated pion can cause
smearing in $\Delta m_{min}$ and dilution of the correlation. 
Selecting a pion with the same charge as the associated pion but with 
different momentum smears events in $\Delta m_{min}$. 
Selecting a pion  with the opposite charge smears events in $\Delta m_{min}$ 
and also dilutes the correlation.

To account for the effects of ghost tracks, smearing, dilution, and acceptance
on the  correlations as functions of
$\Delta m_{min}$, we employ a matrix formalism. For the $D^+$, 
the observed number of  $ D^+ \pi^{\mp} $ combinations 
$O_j^{+\mp}$ in the $j^{th}$ bin of $\Delta m_{min}$ can be written as
\vspace{-3pt}
$$ O_j^{+\mp} = \sum_iS_{ji}^{1\mp}A_i^{+\mp}N_i^{+\mp}+\sum_iS_{ji}^{2\mp}
 A_i^{+\pm}N_i^{+\pm}+G_j^{+\mp}O_j^{+\mp} \eqno(2)$$
\vspace{-3pt}
\noindent where $N_i^{+\mp}$ denotes 
the true number of $D^+ \pi^{\mp} $ events
in the $ i $th bin of $ \Delta m_{min} $, $A_i^{+\mp}$ the 
acceptance probability, and $G_j^{+\mp}$ the ghost track rate for 
$D^+\pi^\mp$ 
combinations. 
The matrix $S^{1\mp}$ describes smearing in the absence of 
dilution while the matrix $S^{2\mp}$ describes smearing and 
dilution when the wrong sign pion is selected.  
The smearing matrices $S^{1\mp}$ and $S^{2\mp}$, the acceptance coefficients 
$A^{+-}$ and $A^{++}$, and the ghost track rates 
$G^{+-}$ and $G^{++}$ are determined 
from the Monte Carlo. The coupled matrix equations in (2) are solved to 
obtain the true distributions $N_i^{+-}$ and $N_i^{++}$.
Corrected $\Delta m_{min}$ distributions are shown in Figure 1. 
The corrected correlation parameters for $D$, $\alpha(D)$, 
for ${\overline D}$, $\alpha({\overline D})$, 
and for the $D$ and ${\overline D}$ combined, $\alpha(D,{\overline D})$ 
are presented in column 4 of Table 1. 

The statistical and systematic errors assigned to the final measurements,
shown first and second respectively, are also given in Table 1. 
These errors are propagated through the matrix formalism.  
The systematic errors account for uncertainties
in the Monte Carlo simulation of the detector (their effects on dilution,
smearing, ghost tracks, and acceptance), analysis cuts, background subtraction,
kaon misidentification, and binning (in decreasing order of
importance as listed). For each data point, the systematic uncertainties due 
to these sources are added in quadrature. The systematic uncertainties due to 
statistical fluctuations in the Monte Carlo are negligible.

To verify the results produced by the matrix formalism, we also estimate
the correlations using simple dilution factors (summed over 
all bins of $\Delta m_{min}$). 
For $D^+$, the true number of combinations,
$N_t^{+-}$ and $N_t^{++}$, can be expressed in terms of the 
reconstructed combinations $N_r^{+-}$ and $N_r^{++}$ as
$$N_r^{+\mp} = (1-d_{+\mp})N_t^{+\mp} + d_{+\pm}N_t^{+\pm}, \eqno(3)$$
\noindent where the dilution factor $d_{+\mp}$ denotes the probability
that a true $D^+\pi^\mp$ combination is reconstructed as a $D^+\pi^\pm$.
The results from this technique are consistent with those reported in Table 1. 

All studies and corrections have been done within the framework of the
LUND PYTHIA/JETSET model.
The dilution factors $ d_{ a b } $ in Eq.~(3) are typically of order
$ 0.2  - 0.3 $.
In our Monte Carlo, $ d_{+-} \approx d_{++} $ but
$ d_{-+} $ is less than $ d_{--} $. 
The difference between $d_{--}$ and $d_{-+}$
is almost independent of $ x_F $ with a typical
value near 0.06. Varying some of the JETSET fragmentation parameters
to reproduce our inclusive $ D^+ / D^- $ production
asymmetries as a function of $ x_F $, as described in ref.\cite{exp:e791},
leads to results consistent
with those in Table 1. A fundamentally different model of hadron production 
might change the differences between the $ d $'s discussed 
above by a few times 0.01, which would in turn
change the measured correlation parameters.
For example, reducing ($ d_{--}  - d_{-+} $) from 0.06 to 0.05 
would increase $ \alpha(D^-) $ by 0.02 -- 0.03.

In Figs. 1(a) and (b) we present the numbers of
$D^0\pi^\pm$ and $\overline {D {}^0}\pi^\mp$ combinations
as functions of $ \Delta m_{min} $. In both of these plots the combinations
differ mainly in the $D^{*\pm}$ resonance region 
,the first 75$\rm MeV/c^2$ bin. 
Using a $ \pm  2.5 \sigma $ cut on the $D^{*+}-D^0$ 
and $D^{*-}-\overline {D {}^0}$ mass 
difference, we separate the final $D^0\pi^+$ and $\overline {D {}^0}\pi^-$ samples  
into resonance ($res$) and continuum ($cont$)
contributions to obtain $\alpha(D^0_{res})= 0.98\pm0.04$ and 
$\alpha({\overline {D {}^0}_{res}}) = 0.98\pm0.02$. 
For pure resonance, $\alpha$ would be near 1. 
The measured values serve as a check of our method.
The continuum measurements are 
$\alpha(D^0_{cont})=-0.07\pm0.03$ and $\alpha(\overline {D {}^0}_{cont})
=0.17\pm0.03 $. 
In Figs. 1(c) and (d) we present the
$D^+\pi^\mp$ and $D^-\pi^\pm$ combinations.
In both these plots the combinations differ over a broad range in 
$\Delta m_{min}$. 
In Figs. 1(e) and (f) we present the $D^{*+}\pi^\mp$ and $D^{*-}\pi^\pm$
combinations. A pattern similar to that for $D^\pm$ is manifest. 
The plots for charm mesons and anticharm mesons clearly differ.
These differences also appear in column (4) of Table 1, and indicate the 
presence of significant beam-related effects.

To investigate beam-related effects in more detail, we study the 
$x_F$ dependence of the $D^{+}$ and $D^{*+}$ correlations. 
We do not show the correlations for $D^0$'s since many $D^0$'s are decay
products of either $D^{*0}$ and $D^{*+}$, making interpretation difficult.
In Fig. 2, we plot $ \alpha $ as a function of $x_F$, 
for both particle and antiparticle for $D^+$ and $D^{*+}$. 
\noindent
The distributions are corrected using the simple dilution factor technique.
We observe that $\alpha(D^+)$ rises slightly with $x_F$ but $\alpha(D^-)$ 
falls sharply to negative values for $x_F > 0.2$.
In both cases, the $ D $'s are more likely to be associated
with $ \pi^- $'s at high $ x_F $ where beam effects seem to be important.
There appears to be less dependence of $\alpha$ on $x_F$ for the $D^{*\pm}$.

Further beam-related studies with Monte Carlo data suggest the correlation 
asymmetries cancel under neutral beam conditions and are in fact symmetric.
This is effectively accomplished when the combined particle and antiparticle 
correlations are computed.
In Table 1, we show the 
combined and symmetrized correlation parameters to be
$\alpha(D^0,\overline {D {}^0}) = 0.29 \pm 0.02 \pm 0.03$,
$\alpha(D^+,D^-) = 0.21 \pm 0.02 \pm 0.03$, and
$\alpha(D^{*+},D^{*-}) = 0.23 \pm 0.04 \pm 0.03$.  
These results indicate that fragmentation dynamics and resonant decays 
produce substantial correlations between  $ D $ mesons and their 
associated pions.
All three combined correlation levels are approximately equal, 
although the correlations for neutral and charged $D$ mesons are dominated by 
resonant and continuum regions of $ \Delta m_{min} $,  respectively.

In summary, we observe significant production correlations between
$ D $ mesons and their associated pions. 
Some of these correlations are associated with fragmentation dynamics,
some with resonances, and some with the charge of the beam.
In addition to providing information on how heavy quarks evolve into hadrons,
such correlations may provide tools for tagging flavor in {\em CP} violation 
studies in heavy flavor systems. 

We gratefully acknowledge the staffs of Fermilab and of all the 
participating institutions. This research was supported by the Brazilian 
Conselho Nacional de Desenvolvimento Cient\'\i fico e Technol\'ogio, 
the Mexican Consejo Nacional de Ciencia y Tecnologica, 
the Israeli Academy of Sciences and Humanities, the U.S. Department 
of Energy, the U.S.-Israeli Binational Science Foundation and the U.S. National 
Science Foundation. Fermilab is operated by the Universities Research 
Association, Inc., under contract with the United States Department of Energy.

\newpage
\begin{table}[btp]
\caption{
The $ x_F $- and $\Delta m_{min} $-integrated correlation parameters 
$\alpha$ defined
in Eq.~(1) for the background-subtracted signals prior to correction,
for the corresponding backgrounds, and for the signals after correction
using the matrix technique based on Eq.~(2). $\alpha( {D {}^0})$
and $\alpha(\overline {D {}^0})$ contain both $D^*$ resonance and nonresonance
contributions.
}  
\label{tab1}
\begin{tabular}{lccc}
Charm & Signal $ \alpha $  & Background $ \alpha $
      & Corrected Signal $ \alpha $ \\
\tableline
$D^0$ & $0.13\pm0.01$ & $-0.04\pm0.01$ & $0.12\pm0.03\pm0.04$ \\
${\overline D^0}$ & $0.18\pm0.01$ & $0.04\pm0.01$ &
$0.42\pm0.02\pm0.03$ \\
$D^0,{\overline D^0}$ & $0.16\pm0.01$ & $0.00\pm0.01$ &
$0.29\pm0.02\pm0.03$ \\
$D^+$ & $0.18\pm0.01$ & $0.10\pm0.01$ & $0.45\pm0.03\pm0.03$ \\
$D^-$ & $0.08\pm0.01$ & $0.02\pm0.01$ & $0.03\pm0.03\pm0.04$ \\
$D^+,D^-$ & $0.13\pm0.01$ & $0.05\pm0.01$ & $0.21\pm0.02\pm0.03$ \\
$D^{*+}$ & $0.15\pm0.02$ & $0.08\pm0.03$ & $0.33\pm0.05\pm0.03$ \\
$D^{*-}$ & $0.08\pm0.02$ & $0.02\pm0.03$ & $0.15\pm0.05\pm0.04$ \\
$D^{*+},D^{*-}$ & $0.12\pm0.01$ & $0.05\pm0.02$ &
$0.23\pm0.04\pm0.03$ \\
\end{tabular}
\end{table}

\newpage
\begin{figure}[htb]
\centerline {\epsfxsize=3.17in \epsffile{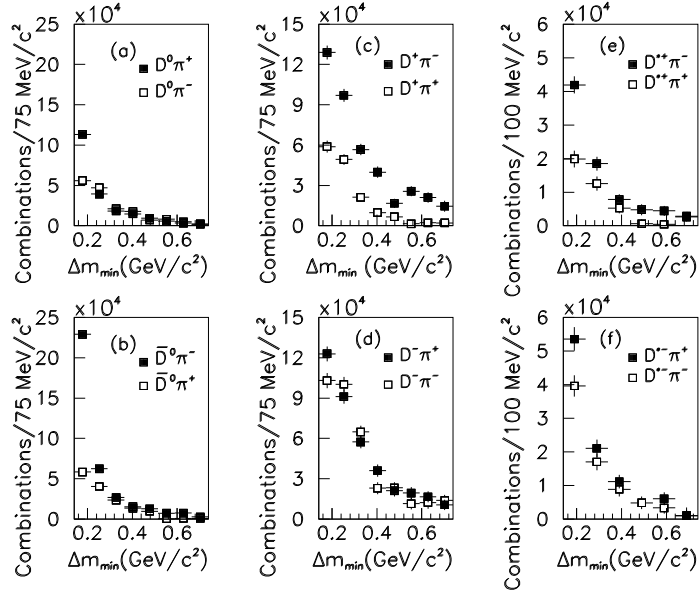} }
\caption{ The fully corrected $ \Delta m_{min} $ distributions for 
(a)$ D^0\pi^\pm $, 
(b)$ \overline {D {}^0}\pi^\mp $, (c)$ D^+\pi^\mp $, (d)$ D^-\pi^\pm $, 
(e)$ D^{*+}\pi^\mp $, and (f)$ D^{*-}\pi^\pm $ combinations.
The error bars are statistical only. } 
\label{fig:ex}
\end{figure}

\newpage
\begin{figure}[htb]
\centerline {\epsfxsize=3.17in \epsffile{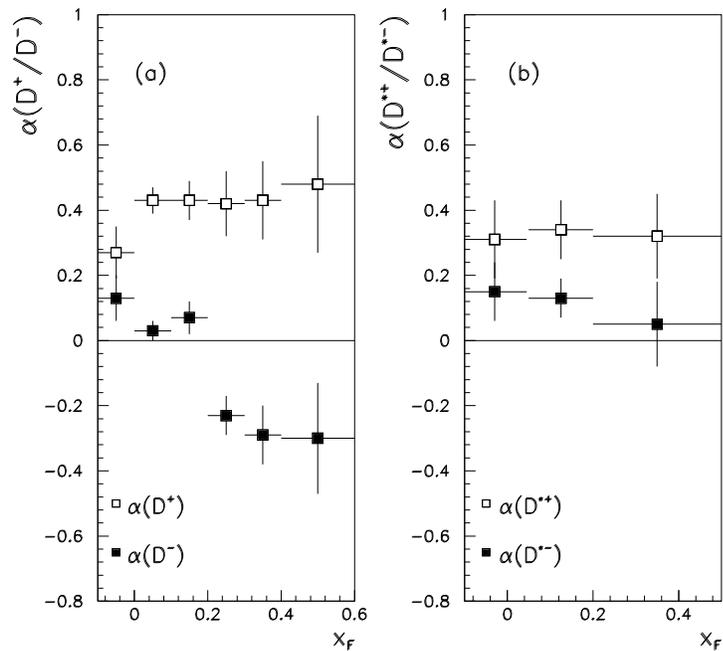} }
\caption{The corrected correlation parameter $\alpha$ as a function of
$x_F$ for (a)$D^+$ and (b)$D^{*+}$.  The parameter $\alpha$ is defined
in Eq. (1) in the text. The error bars correspond to the statistical and 
systematic uncertainties added in quadrature. Additional model-dependence
is discussed in the text.} 
\label{fig:xf}
\end{figure}

\end{document}